\documentclass[12pt,epsf,amssymb]{article} \textheight =23 truecm
\def\lsim{\raise0.3ex\hbox{$<$\kern-0.75em\raise-1.1ex\hbox{$\sim$}}}
\def\gsim{\raise0.3ex\hbox{$>$\kern-0.75em\raise-1.1ex\hbox{$\sim$}}}
\def\noi{\noindent}  \def\bea{\begin{eqnarray}}
\def\eea{\end{eqnarray}} \def\beq{\begin{equation}}
\def\eeq{\end{equation}} 
\def\beeq{\begin{eqnarray}} \def\eeeq{\end{eqnarray}} \def\R{ {\rm R
\kern -.31cm I \kern .15cm}} \def\C{ {\rm C \kern -.15cm \vrule
width.5pt \kern .12cm}} \def\Z{ {\rm Z \kern -.27cm \angle \kern
.02cm}} \def\N{ {\rm N \kern -.26cm \vrule width.4pt \kern .10cm}}
\def\1{{\rm 1\mskip-4.5mu l} }
 
\begin{document} 
\vbox to 1 truecm {}
\centerline{\large\bf  Decoherence and reduction}

\vskip 1 truecm

\centerline{\bf Roland Omn\`es\footnote{e-mail: Roland.Omnes@th.u-psud.fr}}
\centerline{Laboratoire de Physique Th\'eorique\footnote{Unit\'e Mixte de
Recherche (CNRS) UMR 8627}}  \centerline{Universit\'e de Paris XI, B\^atiment
210, F-91405 Orsay Cedex, France}

\vskip 2 truecm

\begin{abstract}
A reduction mechanism resulting directly from the basic principles of quantum mechanics is proposed, inseparably from decoherence. A rather consistent theory of this effect is given and the next problems it raises are indicated. 
\end{abstract}

\vskip 2 truecm

\noindent PCAS codes : 03.65 w, 03.65 Ta, 03.65 Yz 
\vskip 1 truecm

\noindent LPT Orsay 06-24\par
\noindent April 2006\par

\newpage
\pagestyle{plain}
Modern interpretation relies on the idea that basic quantum principles might generate the concepts entering in their own interpretation.  It gave rise to decoherence [1-3], a quantum formulation of classical physics \cite{4r}, a derivation of the Copenhagen rules \cite{5r} and older paradoxes were also removed by means of consistent histories \cite{6r}. The problem of physical reality and its uniqueness remains however controversial and I propose here a new theory for answering this last great question. The main idea relies on considering a hidden part of the state of a large system, which always remains after decoherence and could generate a reduction process without any addition to the quantum principles.\par

I consider first an isolated measuring system consisting of a measured microscopic object and a measuring apparatus. Its Hamiltonian $H = K + E + C$ consists of three parts, $K$ for the collective decohering part, $E$ for the environment, and a coupling $C$. The density matrix $\rho$ obeys the Schr\"odinger equation and its trace $\rho (K)$ over the environment is the reduced density matrix. One can obtain the familiar master equation for $\rho (K)$ in the following way \cite{7r}. One introduces a decohering density $\rho '= \rho(K)  \otimes \rho(E)$, where $\rho (E) = \exp(-\beta E)/Z$ has unit trace and depends on a parameter $\beta$. The hidden part of the density matrix, which will be very important here, is defined by $\rho ''= \rho - \rho '$. Denoting a full trace by $Tr$ and a trace over the environment by $tr$, one subtracts from $C$ a possible collective part acting in the $K$-Hilbert space, so that $tr(\rho C)  \equiv  tr(\rho (E)C) = 0$ (after such a subtraction, the pressure effects of a gas on a piston are included for instance in the collective Hamiltonian $K$). If $A$ is any collective observable in the $K$-Hilbert space, one has then $Tr((A \otimes I(E)) \rho '') = 0$, $I(E)$ being the unit matrix in the environment Hilbert space. As a special case, one has $Tr(\rho '') = 0$. \par

The remaining coupling consists generally of small fluctuations and the master equation for decoherence is obtained from perturbation calculus in $C$, which gives:
\bea
\label{1e} 
&&d\rho (K)/dt = -i[ K, \rho (K)] - i tr([C, \rho '']) \ , \\
&&d\rho ''/dt = - i [K + E,\rho '' ] - i [C, \rho ']\ .			
\label{2e} 
\eea

	\noi Denoting by 0 the initial time at which the measurement begins, one often assumes in decoherence theory that $\rho ''(0) = 0$. I do not make this assumption, so that the solution of Eq. (\ref{2e}) is given by:
\beq
\label{3e}
\rho '' (t) = - i \int_0^t dt' U(t-t') [C, \rho ' (t')]U^{-1} (t-t') + \sigma (t)
\eeq
	
\noi with $U(t) = \exp(-i(K + E)t)$ and $\sigma (t) = U(t)\rho '' (0)U^{-1}(t)$. When the first term in the right-hand side of Eq. (\ref{3e}) is inserted into Eq. (\ref{1e}), one obtains a master equation showing decoherence.  There are as many decoherence channels, as there are different values of the measured observable in the initial state, each channel being characterized by a projection operator $\Pi_j$. The reduced density matrix $\rho (K)$ becomes then rapidly ``diagonal'', i.e. equal to the sum of the channel density matrices $\Pi_j \rho (K)\Pi_j$.\par

	One may notice that the channel projection operators $\Pi_j$ describe mainly classically meaningful properties (asserting for instance together the position and momentum of a pointer). Perhaps more significant and rarely noticed is the fact that decoherence involves non-linear approximations since the parameter $\beta$ is defined through the non-linear equation $\partial \log Z(\beta )/\partial\beta =-Tr(E\rho )$ and $\rho ''(0) = \rho (0) - (tr \rho (0))  \otimes \rho (E)$, where $\rho (E)$ depends on $\beta (0)$. \par

The second step consists in considering the effect of the hidden density $\sigma$ on the probability $p_j = Tr(\Pi_j \rho (K)\Pi_j )$ of channel $j$. This quantity can vary with time according to Eqs.(\ref{1e}-\ref{3e}), at least formally, the corresponding variation being given by 
\beq
d p_j /dt = Tr\left ( \Pi_j [C, \sigma]\right ) \ .				
\label{4e}
\eeq

In the case of a pointer with mass $m$ and center-of-mass position $X$,  interacting with an external atmosphere, the effective coupling $C$ commutes with $X$ and diagonalization occurs in the $|x>$ basis of eigenvectors of $X$ [8]. The operators $C$ and $\Pi_j$ commute and the variation (\ref{4e}) vanishes. But if the pointer slides also along a ruler with mass $m'$, center-of-mass position $X'$, the simplest coupling of this collective subsystem with a phonon with frequency $\omega$ and creation/annihilation operators $a$, $a^{\dagger}$, can be written as  $\lambda A^{\dagger} a + \lambda^* A a^{\dagger}$, with $A = X - X' - (i/\omega )(P/m-P'/m')$. The quantity $\lambda$ measures the strength of the coupling and this expression is obtained for instance when two linear chains, representing the pointer and the ruler, are very close to each other. The coupling $C'$ with a phonon bath does not therefore commute with $\Pi_j$ and the quantity (\ref{4e}) does not vanish. Introducing the eigenfunctions of $\sigma$  and its eigenvalues (positive and negative with vanishing sum), one finds no reason why the variation (\ref{4e}) should vanish. Furthermore, there is no collective basis in which diagonalization would take place when the two decoherence mechanisms act together. Decoherence still acts however, with semi-classical projection operators $\Pi_j$  \cite{7r}, but there is still no evidence why the variation (\ref{4e}) should vanish and it appears then that the probabilities resulting from a thorough analysis of decoherence can vary with time!\\

What does it mean?  An analogy with a priori probabilities may help understanding the situation. A man betting for instance on a horse race relies on a priori estimates of the winning probabilities, on which he makes his bet. During the race, his mind changes however with every event and his mental estimate of probabilities varies accordingly, till the end when certainty is reached and one horse has won with probability 1, the other probabilities having vanished. In the present case, the channel probabilities appear as a priori probability estimates by somebody betting on the validity of quantum theory, which he or she understands through decoherence. \\ 

The reaction of some experts, when they are confronted with this unconventional remark, is that the effect could exist, but would be highly negligible. Is it so, really?  The answer depends obviously on a better knowledge of the density $\sigma$ or, equivalently, of the hidden density $\rho ''(0)$. As a matter of fact, the system that was assumed isolated at the beginning of this discussion is not isolated forever. Even if it is so perfect that the coupling and the channel projections commute exactly, it becomes entangled with the rest of the world as soon as isolation ends (when one looks at the apparatus, for instance) and surely some couplings will then have the destructive effect I mentioned. The question is therefore concerned with what happened before the measurement as much as what will happen afterwards. \\

Every system on the earth is made of particles, which had interactions with many other ones during billions of years and one may reasonably consider the solar system as isolated, at least for the sake of argument. Innumerable decoherence effects occurred within it, although most of them were never amplified to a macroscopic level. A hidden density $R''$ of the solar system was therefore continuously growing, although keeping always a vanishing trace and giving no contribution to the average value of a macroscopic observable.  Its growth is better understood with the example of an isolated system, initially in a pure state with wave function $\psi$. It can be split into a collective part and an environment and undergoes decoherence, but must nevertheless remain in a pure state \cite{9r,10r}. Using the previous notations, its hidden density is $\rho ''  =  |\psi> <\psi| - \rho '$ and the vanishing contribution of $\rho ''$ to the average value of any collective observable implies $Tr(\rho '' \rho ') = 0$, from which one gets $Tr(\rho{ ''}^2) = 1 - Tr(\rho{'}^2)$. In the case of the solar system, one expects therefore $Tr(R{''}^2) \gg Tr(R{'}^2)$ and the initial density  $\rho ''(0)$ of a subsystem is the trace of $R''(0)$ over everything outside the subsystem and has also similar properties. One is therefore led to the puzzling expectation that the change in a priori probability of a channel might not be negligible, because $R''$ is tremendously large in view of the number and perhaps the size of its positive and negative eigenvalues and in spite of its invisibility. The same is probably  true for its partial trace $\rho ''(0)$.\par

When considering the contributions of the eigenfunctions  of $\rho ''(0)$ to the variations $\delta p_j$ during an infinitesimal time $\delta t$, one expects them to generate  a Gaussian Brownian process for the changes in a priori probabilities $p_j$. The sum over $p_j$'s remains however equal to 1, since $\Pi_j$ in Eq. (\ref{4e}) is replaced by the identity operator $I$ in the expression of this sum. The agreement of Born's rule with observation should require however that, during an individual measurement, the Brownian process ends up with one probability $p_j$ being 1 and the other ones equal to 0, {\it the Brownian probability for this outcome being equal to the a priori probability $p_j$, as given by Born's rule}. If this is true, the frequencies of a large number of outcomes will be in agreement with Born's rule. I call ``reduction'' this ``race'' of a priori probabilities during an individual measurement, which ends when one of them reaches certainty.\par

The last step in the analysis consists in checking this expectation. Pearle met a similar problem long ago in a different context \cite{11r} and I investigated it in detail under similar conditions to the present ones, at least from a mathematical standpoint \cite{12r}. The main results were as follows. One considers a finite number $n$ of channels. The evolving quantities $p_j (t)$ are considered as Cartesian coordinates of a moving point $P (t)$ undergoing Brownian motion in a $n$-dimensional Euclidean space $E$, starting from initial values $p_j (0)$ obeying Born's rule (i.e. there is decoherence before reduction). The point $P(t)$ moves in the subspace $E'$ with equation $\sum p_k = 1$ and I suppose the Brownian motion isotropic in $E'$.  I call ``vertex $k$'', or $V_k$, the point in $E$ having all its coordinates equal to 0, except for the $k$-th one, which is equal to 1. The $n$ points $V_k$, are the vertices of a $(n-1)$-dimensional regular polyhedron, or simplex, which is denoted by $S$. When $n = 3$, $S$ is an equilateral triangle. The quantities $p_j (t)$ are the so-called barycentric coordinates of $P(t)$, $p_j (t)$ being its Euclidean distance to the $(n - 2)$-dimensional boundary  simplex opposite to $V_j$ . It is important to notice that every barycentric coordinate is given by a first-degree polynomial in terms of Cartesian coordinates in $E'$, so that it is a harmonic function ($\Delta p_j (t)  = 0)$. \par	

Brownian motion of $P(t)$ must bring it sooner or later onto some face $S_k$ of $S$, on which one of the coordinates, for instance $p_k$, vanishes. One can show that $P(t)$ will never go back into $S$ afterwards, although I cannot give here the too long proof. After reaching such a $(n-2)$-dimensional boundary simplex, $P(t)$ continues its Brownian motion inside it until reaching a $(n - 3)$-dimensional simplex and so on, till it reaches finally a vertex $V_j$. The main problem consists in proving that the Brownian probability $\pi_j$ for reaching $V_j$  is equal to $p_j (0)$, which is itself given by Born's rule.\par

{\it Proof:} One introduces the Brownian probability $q_k$ for $P(t)$ to start from $P(0)$ and to reach the face $S_k$. This function $q_k$ of $\{ p_j(0)\}$ is harmonic in $E'$.  To prove this result, one needs only showing that the value of $q_k$ at the point $P(0)$ is the average of $q_k(Q)$ over the points $Q$ lying on a $(n - 2)$-dimensional sphere $\Sigma$ centered at $P(0)$ and inside $S$. Writing $P$ in place of $P(0)$ and denoting by $dw$ an element of area on $\Sigma$ around $Q$, the harmonic character of $q_k$ will be established if one has
\beq
\label{5e}
q_k(P) = \Omega^{-1} \int_{\Sigma} q_k (Q) d\Omega\ .
\eeq

\noi But after starting from $P(0)$, the point $P(t)$ must cross $\Sigma$ before reaching $S_k$. Letting $\Pi (P, Q) d\Omega$  denote the probability for reaching $\Sigma$  for the first time in the region $d\Omega$  containing $Q$ and using composite probabilities, one has:
\beq
\label{6e}
q_k(P) = \int_{\Sigma} \Pi(P, Q) q_k (Q) d\Omega\ .
\eeq

\noi But $\Pi (P, Q) = \Omega^{-1}$ since the Brownian motion is isotropic, so that Eq. (\ref{6e}) must be true.

When the descending motion of $P(t)$ brings it to its last but one step, it has reached a one-dimensional simplex, which is an interval such as $V_1V_2$, at some point with coordinate $p_1$. The previous result shows that the probability for reaching finally the vertex $V_1$ is a harmonic function of $p_1$ (i.e. a first-degree polynomial), which is obviously equal to 1 at $V_1$ and zero at $V_2$. It is therefore equal to $p_1$ and the expected theorem is thus proved for $n = 2$. One can then climb up the previous descent to prove the generality of this result. When $n = 3$, one already knows from the case $n = 2$ that $\pi_1$ is equal to $p_1(0)$ on both sides $V_1V_2$ and  $V_1V_3$ of the triangle $S$ and zero on the side $V_2V_3$ (where  $p_1(0) = 0$). But the function identically equal to $p_1 (0)$ is harmonic and satisfies the boundary conditions. Therefore $\pi_1 =  p_1(0)$ and more generally $\pi_k =  p_k(0)$.It can also be shown that this result cannot hold if the Brownian motion is non-isotropic or non-homogeneous, a non-homogeneous motion being one in which the Brownian correlations depend on the location of $P(t)$ \cite{12r}. Elementary considerations on the Brownian variation of a quantity $p_j$ suggest also that the time of reduction does nor depend much on the number of channels.\\

Reduction is certainly not a local effect, as shown by its necessary homogeneity and by conservation of the sum of a priori probabilities over all channels. This essential character is not surprising from an intuitive standpoint in which $\sigma$ represents phase correlations,  accumulating and always made more complex in the course of time. The eigenfunctions of $\sigma$ are therefore not localized. One can show that this implies also Brownian variations in the non-diagonal cross-channels $\Pi_j\rho (K)\Pi_k$ ($j \not= k$), but this effect is certainly very rapidly wiped out by decoherence.\par

My conclusion will be that the effect exists, because it follows directly from the basic quantum principles. It would be ineffective only if something important had not yet been understood in decoherence theory (for instance a universal restriction on coupling with the environment) or if the reduction time were extremely large. Unfortunately, this time is unknown empirically since the existence  of a unique reality is {\it consistent} with decoherence \cite{13r}, without reduction. Investigation of the hidden density $\sigma$ is a fascinating and difficult problem, particularly when trying to estimate the reduction time. Models will perhaps give hints, but an interesting approach would be to consider the real state of the vacuum as not being a pure state $|0>$ but rather $|0><0| + \sigma$ (at the field level) and to look for consequences in quantum field theory. Cosmological considerations could also be valuable. The philosophical background of realism will be drastically changed of course if the present theory is valid, but jumping at conclusions on this point would be ill-advised. 
			
\section*{Acknowledgements}
\hspace*{\parindent}
A private discussion with Murray Gell-Mann led me years ago to wonder about hidden density matrices. When proposing the decoherence theory in Reference 12, I benefited from a suggestion by Roger Balian, which proved decisive here. I received also very helpful comments and criticisms when I tried to build up the present theory, for which I thank particularly H-Dieter Zeh and Bernard d'Espagnat .

\end{document}